\documentclass[preprint]{aastex}
\usepackage{graphicx}
\begin{document}
\title{ Numerical simulation of oscillating magnetospheres 
with resistive electrodynamics }%
\author{Yasufumi Kojima$^*$ and Yugo E. Kato }
\affil{Department of Physics, Hiroshima University, 
Higashi-Hiroshima, Hiroshima 739-8526, Japan}
\email{$^*$ ykojima-phys@hiroshima-u.ac.jp}
\begin{abstract}
  We present a model of the magnetosphere around an oscillating neutron 
star. The electromagnetic fields are numerically solved by modeling 
electric charge and current induced by the stellar torsional mode, with 
particular emphasis on outgoing radiation passing through the magnetosphere.
The current is modeled using Ohm's law, whereby 
an increase in conductivity results in an increase in the induced current.
As a result, the fields are drastically modified,
and energy flux is thereby enhanced. This behavior is however 
localized in the vicinity of the surface since the induced current
disappears outwardly in our model, in which the exterior is assumed to 
gradually approach a vacuum.
\end{abstract}

\section{Introduction}
  Giant flares in Soft Gamma-ray Repeaters (SGRs) are 
some of the most energetic astrophysical events. 
The total energy released 
during a flare is $\sim 10^{46}$ ergs, and flares are considered to be
powered by a global reconfiguration of magnetic 
field $> 10^{14}$ G in a magnetar (\cite{2008A&ARv..15..225M} 
for a review).
Global seismic oscillations of the stellar crust
are likely to be associated with flares\citep{1998ApJ...498L..45D}.
Quasi-Periodic Oscillations (QPOs) are observed in the decaying
tails of the flares of SGR 1900+14 and SGR 1806-20, and 
their frequencies are in the range of 10-$10^{3}$ Hz,
which are interpreted as the frequencies of torsional oscillations with some
harmonic numbers (e.g., \cite{2005ApJ...628L..53I, 2006ApJ...653..593S}).
The typical frequency of the torsional shear mode is $\sim 10$Hz, which is 
estimated from the shear modulus, density at the crust and stellar radius. 
More realistic treatment is necessary for the actual fitting.  
In particular, considering the effects of strong magnetic fields, 
which couple core and crust, is essential.
A number of researchers have been exploring the oscillations of 
strongly magnetized stars
(\cite{2005ApJ...634L.153P,2007MNRAS.377..159L,2008MNRAS.385.2069L,
2008MNRAS.385L...5S,2009MNRAS.397.1607C,2011MNRAS.414.3014C,
2012MNRAS.420.3035V,2012MNRAS.421.2054G} and references therein).
%

 In spite of substantial improvements in 
oscillation theory, there is considerable uncertainty about 
the processes behind X- and gamma-ray radiation phenomena.
Main peaks in the power spectrum may correspond
to the oscillation frequency, but their amplitudes 
and decays should be affected by the surrounding magnetosphere.
\citet{2000MNRAS.316..734T} showed that an oscillating magnetized star
should have a magnetosphere filled with charge-separated plasma
so that the longitudinal electric field (parallel to the magnetic field)
in vacuum is screened\footnote{
This has been extended to a model in curved spacetime
\citep{2009MNRAS.395..443A,2010MNRAS.408..490M}.
In this paper, the effects of general relativity are ignored.
}.
This situation resembles that of a pulsar magnetosphere. Rotation
in pulsars is replaced by oscillation in magnetized stars. 
The induced charge density has been calculated as a 
generalization of the well-known Goldreich-Julian density 
for pulsars \citep{1969ApJ...157..869G}.
In the almost analytic treatment in \citet{2000MNRAS.316..734T},
the current is assumed to be zero everywhere.
In magnetar flares, large current associated with a 
magnetic twist $ \vec{\nabla} \times \vec{B} \ne 0 $
is expected to flow out from the surface.
Therefore, it is necessary to consider the effects 
of flowing current on the magnetosphere. 
The calculations necessary to consistently determine
the charge and current densities induced by the oscillation are 
more complex, requiring time-dependent numerical simulations.
One approach is to perform relativistic magnetohydrodynamics (MHD) 
simulations. 
However, this approach is rather difficult in the case of strong magnetic
regimes, such as that in magnetars, since plasma effects are less important.
Recently, a new numerical approach was proposed for the construction of 
pulsar magnetospheres\citep{2011arXiv1101.3100G,
2012ApJ...746...60L,2012ApJ...749....2K}.
In this approach, the Maxwell equations are solved for the current modeled 
using Ohm's law, and particle acceleration as well as the generation 
of radiation are relevant to the electric field component parallel 
to the magnetic field, that is, $\vec{B} \cdot \vec{E} \ne 0$.
Thus, such dissipative effects of non-ideal 
MHD are taken into account in the model,
which differs from the force-free 
electrodynamics approach, in which plasma can be approximated
as massless and ideal MHD conditions are assumed to hold everywhere.
%

  We adopt the same resistive electrodynamics approach
in order to simulate global structures in the flares.
The electromagnetic fields outside a star are disturbed by 
periodic torsional motion, which is horizontal at the surface.
There is no justification for the use of resistive electrodynamics
since the portion of plasma energy with respect to electromagnetic energy
ejected from the surface is unknown. 
Moreover, the physical conditions relevant to QPO phenomena of the 
flares is uncertain. Non-thermal X- and gamma-rays in 
pulsars originate from the effects of non-ideal MHD, 
namely accelerating electric fields.
It is therefore a reasonable first step to simulate the disturbances 
in the magnetosphere by analogy with pulsars, in whose framework 
non-ideal MHD is allowed. The results are expected to provide useful insight
into observed phenomena and to lead to further improvement of 
theoretical models.
This paper is organized as follows. In Section 2, we discuss the 
proposed numerical
method for the evolution of electromagnetic fields in two-dimensional
axially symmetric space. We adopt spherical 
coordinates $(r, \theta, \phi)$ and assume axial 
symmetry ($\partial /\partial \phi =0$).
However, in general, the azimuthal component of a vector is nonzero.
For example, a toroidal magnetic field $B_\phi$ can be produced 
by a poloidal current $(j_r, j_\theta )$.
Disturbances associated with shear oscillations are assumed 
at the inner boundary. In Section 3, we present the results of a numerical 
simulation based on the proposed model, and Section 4 presents a discussion 
of the results.
%

\section{Assumptions and Equations}
  \subsection{ Electromagnetic fields}
  Maxwell's equations are solved by assuming a charge 
density $\rho _{e}$ and a current density $\vec{j}$:
\begin{equation}
\frac{1}{c}\frac{\partial \vec{B}}{\partial t}=
  -\vec{\nabla}\times \vec{E},
\label{eqn.Farad}
\end{equation}
\begin{equation}
\frac{1}{c}\frac{\partial \vec{E}}{\partial t}=
  \vec{\nabla}\times \vec{B}-\frac{4\pi \vec{j}}{c},
\label{eqn.Amp}
\end{equation}
\begin{equation}
\vec{\nabla}\cdot \vec{B}=0, 
\label{eqn.Gauss}
\end{equation}
\begin{equation}
\vec{\nabla}\cdot \vec{E}=4\pi \rho _{e}.
\label{eqn.Coulb}
\end{equation}
To solve these equations, we use a scalar 
potential $\Phi $ and a vector potential $\vec{A}$ satisfying 
the Coulomb gauge, $\vec{\nabla}\cdot \vec{A}=0$. 
For axially symmetric fields, the following form given by
two functions, $F(t,r,\theta )$ and $G(t,r,\theta )$, is automatically
satisfied with the gauge condition: 
\begin{equation}
\vec{A}=\frac{1}{r\sin \theta }\vec{\nabla}F\times \vec{e}_{\phi }
+\left( \frac{G}{r\sin \theta }\right) \vec{e}_{\phi },
\end{equation}
where $\vec{e}_{\phi }$ is a unit vector in the azimuthal direction. 
One of the advantages in the potential formalism is that the constraint 
Eq.(\ref{eqn.Gauss})
is automatically satisfied. 
Helmholtz's vector decomposition due to the divergence and rotation 
of $\vec{j}$ is also clear for these functions as discussed below.
The electric field $\vec{E}$ and magnetic field $\vec{B}$ are given by 
the derivatives of the potentials: 
\begin{equation}
\vec{E} =-\vec{\nabla}\Phi -\frac{1}{r\sin \theta }\vec{\nabla}
\left( \frac{\partial F}{c\partial t}\right)\times \vec{e}_{\phi }
-\left( \frac{1}{r\sin \theta }
\frac{\partial G}{c\partial t}\right) \vec{e}_{\phi },  
\label{eqn.E}
\end{equation}
\begin{equation}
\vec{B} =\frac{1}{r\sin \theta }\vec{\nabla}G\times \vec{e}_{\phi }+\left( 
\frac{S}{r\sin \theta }\right) \vec{e}_{\phi }.
  \label{eqn.B}
\end{equation}
The function $S$ in $B_{\phi }$ is given by
\begin{equation}
  S=-{\mathcal{D}}F, 
 \label{eqn.F}
\end{equation}
where the differential operator $\mathcal{D}$ is defined as 
\begin{equation}
{\mathcal{D}}\equiv \left( \frac{\partial ^{2}}{\partial r^{2}}
+\frac{\sin\theta }{r^{2}}\frac{\partial }{\partial \theta }
\left( \frac{1}{\sin \theta }\frac{\partial }{\partial \theta }\right)
 \right) .
\end{equation}
The electric field (Eq. (\ref{eqn.E})) given 
by the time derivative of $F$ is used 
 to facilitate solving the time derivative of Eq. (\ref{eqn.F}),
i.e., 
\begin{equation}
 {\mathcal{D}}
\left( \frac{\partial F}{\partial t} \right) 
=-\frac{\partial S}{\partial t} .
  \label{eqn.FTM}
\end{equation}%
Substituting these forms (Eqs. (\ref{eqn.E}) and (\ref{eqn.B}))
into Maxwell's equations, we obtain two wave
equations for $G$ and $S$, and a Poisson equation for $\Phi$: 
\begin{equation}
 \left( \frac{1}{c^2}\frac{\partial ^2}{\partial t^2 } - {\mathcal{D}}
  \right) G = \frac{4 \pi }{c} j_\phi r \sin \theta,
  \label{eqn.G} 
\end{equation}
\begin{equation}
 \left( \frac{1}{c^2}\frac{\partial ^2}{\partial t^2 } - {\mathcal{D}}
\right ) S = 
\frac{4 \pi}{c} \left( \frac{\partial ( r j_\theta) }{\partial r }
 - \frac{\partial j_r }{\partial \theta } \right) \sin \theta ,
  \label{eqn.S}
\end{equation}
\begin{equation}
 \left( \frac{1}{r^2} \frac{ \partial }{\partial r } \left( r^2 \frac{
\partial }{\partial r }\right) +\frac{1}{r^2 \sin \theta } \frac{ \partial }{%
\partial \theta } \left( \sin \theta \frac{ \partial }{\partial \theta }
\right) \right) \Phi = - 4 \pi \rho_e . 
 \label{eqn.P}
\end{equation}

  The evolution of these functions is based on a model of 
current density. We briefly introduce the model presented in 
\cite{2012ApJ...746...60L}.
Ohm's law relating the current and the electric field is
\begin{equation}
 \vec{j}_{fluid} =  \sigma \vec{E}_{fluid},   
\label{eqn.Ohm}
\end{equation}
where $\sigma $ denotes electric conductivity and  
Eq. (\ref{eqn.Ohm}) is a representation in the fluid rest frame.
Upon performing transformation into quantities in some inertial frame, 
the general form of the current model is given by 
\begin{equation}
 \vec{j} =\rho _{e}\vec{v}_{D}+\Gamma ( B_{0}\vec{B}+E_{0} \vec{E} ),
\label{eqn.CJ}
\end{equation}
where $\vec{v}_{D}$ is generalized 
`$\vec{E}\times \vec{B}$ drift velocity'
\begin{equation}
 \vec{v}_{D}=\frac{\vec{E}\times \vec{B}}{B^{2}+E_{0}^{2}}c,
\end{equation}%
and $E_{0}, B_{0}$ are given by the two Lorentz invariants
$E_{0} ^{2} -B_{0} ^{2} = |\vec{E}|^{2} -|\vec{B}|^{2}$,
$E_{0} B_{0} = \vec{E} \cdot \vec{B}$, and $E_{0} \ge 0. $
The drift velocity contains the term $ E_{0} ^{2} $
in the denominator to account for the non-zero 
electric field in the fluid rest frame.
This correction ensures $| \vec{v}_{D}| \le c$, 
even if the magnetic field vanishes.
The term $\Gamma$ in Eq. (\ref{eqn.CJ}) represents the effects of non-ideal 
MHD. Various model prescriptions are possible since there is 
ambiguity concerning the plasma velocity, in which frame Ohm's law holds.
For a choice of `minimal velocity',\footnote{
The velocity along the magnetic field is zero in the lab frame.}
$\Gamma$ is given by\citep{2012ApJ...746...60L}, 
\begin{equation}
 \Gamma = 
\frac{\sigma E_{0} }{(B_{0}^{2}+E_{0}^{2})^{1/2}(B^{2}+E_{0}^{2})^{1/2}}.
 \label{eqn.LST}
\end{equation}%
For comparison, the expression $\Gamma_{G}$ proposed by 
\cite{2011arXiv1101.3100G} is
\begin{equation}
 \Gamma_{G} = \Gamma \times
\left( 1 + \frac{(\rho _{e} c )^{2}}{ (\sigma E_{0})^{2} }
\frac{B_{0}^{2}+E_{0}^{2}}{B^{2}+E_{0}^{2}}
\right)^{1/2}  ,  
  \label{eqn.Gru}
\end{equation}%
where the charge density is assumed to vanish in the fluid rest frame. 
In addition, note that $ \Gamma $ is not necessarily 
positive, and it is possible to model the energy transfer to
electromagnetic fields in the region where $ \Gamma <0$. 
Thus, at present there is no unique form for non-ideal dissipation. 
We adopt Eq. (\ref{eqn.LST}), which is the simplest form, and 
easily compare our results with those in the pulsar magnetosphere  
by\citep{2012ApJ...746...60L}. 
In more general cases, the plasma velocity should vary temporally 
and spatially, and would provide some complicated structure.
Obtaining a realistic form of $\sigma$ strongly depends on the spatial position.
The conductivity is likely to decrease
to zero with the radius $r$ since the plasma number density decreases. 
To derive the behavior toward the vacuum region, 
we use the simple power law form
\begin{equation}
  \sigma =\sigma _{0} ( r/r_{0} ) ^{-1}
 \label{eqn.modelsg}
\end{equation}
where $\sigma_{0} $ is a constant and $r_{0}$ is the inner boundary radius.

  Finally, the time evolution of $\rho _{e}$
is given by the charge density conservation law
\begin{equation}
\frac{\partial \rho _{e}}{\partial t}=-\vec{\nabla}\cdot \vec{j}.
 \label{eqn.chg}
\end{equation}%
The numerical procedure for calculating the electromagnetic fields 
is summarized here. Once the charge density $\rho _{e}$ and current 
density $\vec{j}$ are given at a particular time, the time evolutions
of $G$, $S$ and $\rho_{e}$ are found by solving 
Eqs. (\ref{eqn.G}),(\ref{eqn.S}) and (\ref{eqn.chg})
with appropriate boundary conditions.
The evolution of these functions is separately governed 
by the nature of vector 
field $\vec{j}$. More specifically,
$G$ is governed by the toroidal component $j_{\phi}$,
$S$ is governed by the rotation of poloidal components 
$\vec{\nabla} \times \vec{j}_{p}$, and
$\rho_{e}$ is governed by their divergence  
$ \vec{\nabla} \cdot \vec{j}_{p}$.
The elliptic equation (\ref{eqn.P}) is solved at each time step
for the function $\Phi $ with $\rho_{e}$. 
There is another elliptic equation (\ref{eqn.FTM})
for the function $\partial F/\partial t$.
Thus, the electromagnetic fields are given by Eqs. (\ref{eqn.E}) 
and (\ref{eqn.B})
in terms of the functions $G$, $S$, $\Phi$ and $\partial F/\partial t$,
and the current $\vec{j}$ is also updated by Eq. (\ref{eqn.CJ}).

   \subsection{ Torsional oscillation}
  We consider disturbances in the electromagnetic fields
in the presence of torsional oscillation within a star.
Before the onset, the exterior fields are in the form of a purely magnetic 
dipole given by
\begin{equation}
\vec{B}_{d} = [B_{r}, B_{\theta }, B_{\phi } ] 
 = \left [ \frac{2 \mu \cos \theta }{r^3}, 
           \frac{ \mu \sin \theta }{r^3},  0  \right ],      
\end{equation}
where $\mu$ is the dipole moment and the typical field strength $B_{s} $
at the surface $r_{0} $ is $B_{s}=\mu /r_{0}^3 $.
The magnetic flux function is 
\begin{equation}
 G = \frac{\mu \sin ^2 \theta }{r} .
  \label{eqn.puredip}
\end{equation}

  We examine the axially symmetric torsional oscillation,
whose motion is horizontal in the azimuthal direction
at the surface.
In general, the velocity associated with the sinusoidal oscillation 
with angular frequency $ \omega $ is given by
\citep{1989nos..book.....U}
\begin{equation}
v _{\phi } = -\epsilon c \sin( \omega t)
\frac{ \partial P_{l} }{ \partial \theta} ,
  \label{eqn.vlpert}
\end{equation}%
where $ P_{l}(\theta)  $ is a Legendre function of order $l$,
and $\epsilon $ denotes the dimensionless amplitude of the oscillation. 
The frozen-in condition, $\vec{E}+\vec{v}\times \vec{B}/c=0$, 
results in a poloidal electric field $\vec{E}_{p}$ at the surface:
\begin{equation}
\vec{E}_{p} = [E_{r}^{(0)} , E_{\theta }^{(0)} ] = 
  \left[
 -\epsilon B_{s} \sin( \omega t)  \sin \theta  
  \frac{ \partial P_{l} }{\partial \theta} , 
   2\epsilon B_{s} \sin( \omega t)  
 \cos \theta \frac{ \partial P_{l} }{\partial \theta} 
\right].
 \label{eqn.Ep0}
\end{equation}%
In the exterior vacuum region, 
the torsional mode with $l$ in general induces 
outgoing radiation with $l \pm 1$\citep{1984ApJ...281..746M}.
Below, we consider oscillation with $l=2$ only
in order to examine dipole radiation, 
in which case the damping is most effective. 
The azimuthal component $E_{\phi }$ is not induced since it belongs to 
a different type of mode. Electromagnetic wave solutions in vacuum are
classified into two modes: the TE mode $(\vec{E}_{p}, B_{\phi })$ 
and the TM mode $(\vec{B}_{p}, E_{\phi })$\citep{1975clel.book.....J}.  
The latter is described solely by the function $G$,
which is given by the continuity of $B_{r}$.
That is, $G$ at $r_{0}$ is always fixed as a dipolar 
function (Eq. (\ref{eqn.puredip})).
The azimuthal component of the electric field ($E_{\phi }=0 $)
also satisfies the continuity across the surface.
On the other hand,  $B_{\theta}$ is not specified at the surface, 
since it is described by the derivative 
$ \partial G/\partial r$, and it is not in general continuous
due to the surface currents.

 We discuss the boundary conditions of TE part $(\vec{E}_{p}, B_{\phi })$. 
The tangential component $E_{\theta} $ should be continuous.
Using $E_{\theta} ^{(0)}$ ( Eq. (\ref{eqn.Ep0}) ) for $l=2$, 
the explicit forms of $\Phi $ and $F$ 
in Eq. (\ref{eqn.E})  can be chosen at $r_{0}$ as
\begin{equation}
\Phi = -2 \epsilon r_{0} B_{s} \sin(\omega t) \cos ^{3} \theta ,
~~~
\frac{\partial F}{\partial r} =0.
\end{equation}
The radial component $ E_{r} $ is not continuous 
to $E_{r}^{(0)}$ ( Eq.(\ref{eqn.Ep0}) ) 
due to the surface charges.
Remaining boundary condition for $B_{\phi } (= S/(r\sin \theta))$
is specified by a physical argument.
The derivative $ \partial B_{\phi }/\partial r$
is related with $ \partial E_{\theta }^{(0)}/\partial t$
and $j_{\theta} $ in Eq.(\ref{eqn.Amp}). 
By assuming that the tangential component $j_{\theta} $ vanishes, 
the explicit form of $S$ at $r_{0}$ can be described by
\begin{equation}  
\frac{\partial S}{\partial r} =
\frac{6 \epsilon \omega  r_{0} B_{s}}{c} 
\cos(\omega t) \cos^{2}  \theta \sin ^{2} \theta .
  \label{eqn.BCBax}
\end{equation}%
The assumption  $j_{\theta} =0$ is good at least in the polar regions, 
since the currents flow along the magnetic field, which 
is radial near the poles.
The boundary condition (\ref{eqn.BCBax}) is not a
unique one, since there exist a number of models for the out-flowing 
currents at $r_{0}$. 
In exterior vacuum, i.e., $\vec{j}=0$ everywhere,
the condition (Eq. (\ref{eqn.BCBax})) provides some amount
of outgoing energy power, which is analytically calculated.
The model with Eq.(\ref{eqn.BCBax}) is therefore useful 
for comparison.
We consider the response of the 
magnetosphere to the disturbances, which are caused with
the same boundary value as that for the vacuum.
In this paper, we discuss
how and where the ideal MHD condition is broken 
and the resistive effect works in the magnetosphere.

Here, we estimate the outgoing energy flux 
for the boundary condition (Eq. (\ref{eqn.BCBax})). 
The time-averaged Poynting flux at the wave zone 
contains dipole ($l=1$) and octupole ($l=3$) radiation.
By matching $B_{\phi}$ at $r_{0}$ with the 
wave solution of $l=1$\citep{1975clel.book.....J,1984ApJ...281..746M},
we obtain the luminosity as
\begin{equation}  
 L = \frac{12}{25}  (\epsilon r_{0} B_{s})^{2} c
\frac{x_{0}^{4}}{x_{0}^{4}-x_{0}^{2}+1} ,
 \label{eqn.TYP}
\end{equation}%
where $x_{0}=\omega r_{0}/c$. At the low frequency limit, this is reduced to 
$ L \propto  (\omega r_{0}/c )^{4}(\epsilon r_{0} B_{s})^{2} c$.
The dependence on $\omega$ is due to dipole radiation.
Similarly, the luminosity of octupole radiation is of the order of
$   (\omega r_{0}/c )^{8}(\epsilon r_{0} B_{s})^{2} c$,
and is negligibly small for $\omega r_{0}/c \le 1$.

  \subsection{ Numerical method and model parameters}
   We adopt the finite difference method in order to solve the 
partial differential equations.
The numerical domain of the spherical coordinates 
is $1 \le r/r_{0} \le 150 $ and $0 \le \theta \le \pi $. 
The typical number of cells on the grid is $300$ and $180$ in 
the directions of $r$ and $\theta$, respectively.
The grid cell spacing in the radial direction is taken 
as $ \Delta r \propto r^{1/2}$
to obtain fine resolution near the inner region, 
whereas the spacing in the angular direction
is constant.
The boundary conditions at the inner boundary are already discussed
in the previous subsection. The other conditions are regularity conditions
at the polar axis ($\theta =0, \pi$)
and outgoing wave conditions at the outer boundary.
The initial conditions assume a pure dipole, that is, the function $G$ is 
given by Eq. (\ref{eqn.puredip}), and the other functions are zero:
$S=$ $\Phi=$ $\partial F/\partial t =$ $\rho_{e}=$ $\vec{j}=0$.

In the numerical simulations, the oscillation frequency and 
amplitude are chosen to be larger than realistic values.
The QPO frequencies observed in magnetar flares are of the order of
$\nu = 10^{1}$-$10^{3}$Hz.
In particular, 30 Hz QPOs can be interpreted as nodeless shear 
oscillations with $l=2$\citep{2006ApJ...653..593S}, 
for which the frequency is written as
$\nu \sim  10^{-3}c/R_{s}$ 
using the stellar radius $R_{s}$ and $c$.
In our numerical model, the frequency is chosen as
$\nu = \omega/(2 \pi)=0.1c/r_{0}$
(wavelength $\lambda= 10 r_{0}$)
in order to reduce the model size and facilitate the evaluation of 
the luminosity, which is typically given by Eq. (\ref{eqn.TYP}).
The frequency is scaled up by a factor of $10^{2}$
if the inner radius $r_{0}$ is taken as the stellar surface 
radius $R_{s}$.
The scaling of $\nu$ does not hold exactly since the conductivity 
depends on the spatial position. 
The dynamics of our simple model is governed by a dimensionless 
parameter (the ratio of electric conductivity to the frequency), 
as described below.
In future research, it would be necessary to study the dynamics by devising a
more realistic conductivity model.
The oscillation amplitude is also slightly higher, where $\epsilon = 0.1$ 
is used in the numerical calculations, whereas the
actual amplitude of flares is $\epsilon \ll 1$
in Eq. (\ref{eqn.vlpert}).
The scaling of $\epsilon $ is confirmed up to  
$\epsilon \le {\mathcal O}(1)$.

The electric conductivity $\sigma$ is key
to determining the electromagnetic field structure.
For example, in the case of $\sigma =0 $ everywhere, 
the electric current always vanishes 
when the initial conditions assume a purely magnetic dipole, and hence
the state is stationary. The magnitude $ \sigma_{0}$ in the form of the power 
law in Eq. (\ref{eqn.modelsg})
is normalized as $p \equiv \sigma _{0} (4\pi/c) (c/\omega)$
$= 2 \sigma _{0} / \nu$,
where $4\pi/c \approx 377{\rm ohms}$ is 
the characteristic impedance of vacuum\footnote{
This normalization differs by a factor of $4 \pi $ 
from that of a pulsar magnetosphere\citep{2012ApJ...746...60L},
for which the parameter $ \sigma /\Omega $ 
is used with the angular rotation $ \Omega $.
}.
The dimensionless parameter $p= 2 \sigma _{0} /\nu$
denotes the ratio of the current to the displacement current:
$p= 4\pi j /( \omega E) $ $= 4\pi \sigma _{0} / \omega $. 
If $p \gg 1 $, the displacement current is negligible,
whereas the limit of $p= 0 $ corresponds to vacuum.
In magnetar flares, a sufficient amount of 
electron-positron pair plasma may be produced, which  
fills the magnetosphere\citep{2007ApJ...657..967B}.
Therefore, the case of extremely small $p$ is not applicable, although 
it is still useful for comparison.
%

  \section{ Results }
  Figure 1 shows the luminosity, which is the radial component of the 
Poynting flux through a surface. From Eq. (\ref{eqn.TYP}), for 
our numerical values ($\nu r_{0}/c= \epsilon = 0.1$), the typical value
of the luminosity is $L/(  ( r_{0} B_{s})^{2} c)\sim 10^{-3}$ .
The luminosities for $\sigma_{0}/\nu =5$(top 
panel) and $0.5$ (bottom panel) are shown. 
They are estimated at two different radii 
($ r= \lambda (=10 r_{0}) $ and $4\lambda $).
Outgoing waves are modified in the presence of resistivity.
Similar sinusoidal curves with a period of $5 r_{0}/c$
can be seen in the figure\footnote{
Note that the Poynting flux is the square of oscillating quantities,
so that the frequency is doubled ($2 \nu c/ r_{0}=1/5$).
}.
The amplitudes of the waves depend on both the estimated radius and 
the parameter $\sigma_{0}$. Regarding the flux at $r= \lambda $,  
the oscillating amplitude for higher conductivity 
is slightly larger than that for lower conductivity.
The amplitudes decrease with the radius.
Large current is temporarily
induced within the inner region and tends to zero in our 
model with $\sigma \propto r^{-1}$.
The current significantly modifies the electromagnetic
fields and allows for a greater Poynting flux.
With the increase of the parameter $\sigma_{0}$, 
the induced current also increases.
The curves in Fig. 1 are almost completely described by a single Fourier 
component, and are therefore different from the X-ray light 
curve corresponding to flares.
The complexity in the observed power spectrum, such as broad peak width and
rapid time variation in the QPO, is likely to reflect the nonlinearity
of the dynamical system and/or the emission mechanism.
In the numerical simulation, a more complex profile can be produced,
for example, by further increasing the current.
The parameter range $\sigma_{0} /\nu > 10$ is more interesting. 
Actually, we tried and found that the fluctuations in the small scale 
are easily excited, and sometimes grow. This local instability may originate 
from our numerical scheme or the current model.
Performing a comparison also becomes difficult since care should be 
taken to ensure the validity and efficiency of analysis in the case 
of highly complex and time-dependent data. This regime is beyond 
the scope of this paper,
and the parameters are limited to certain values 
$\sigma_{0} /\nu =p/2 \le {\mathcal O}(10)$.

\begin{figure}[!tb]
%
  \includegraphics[scale=1.2, angle=0]{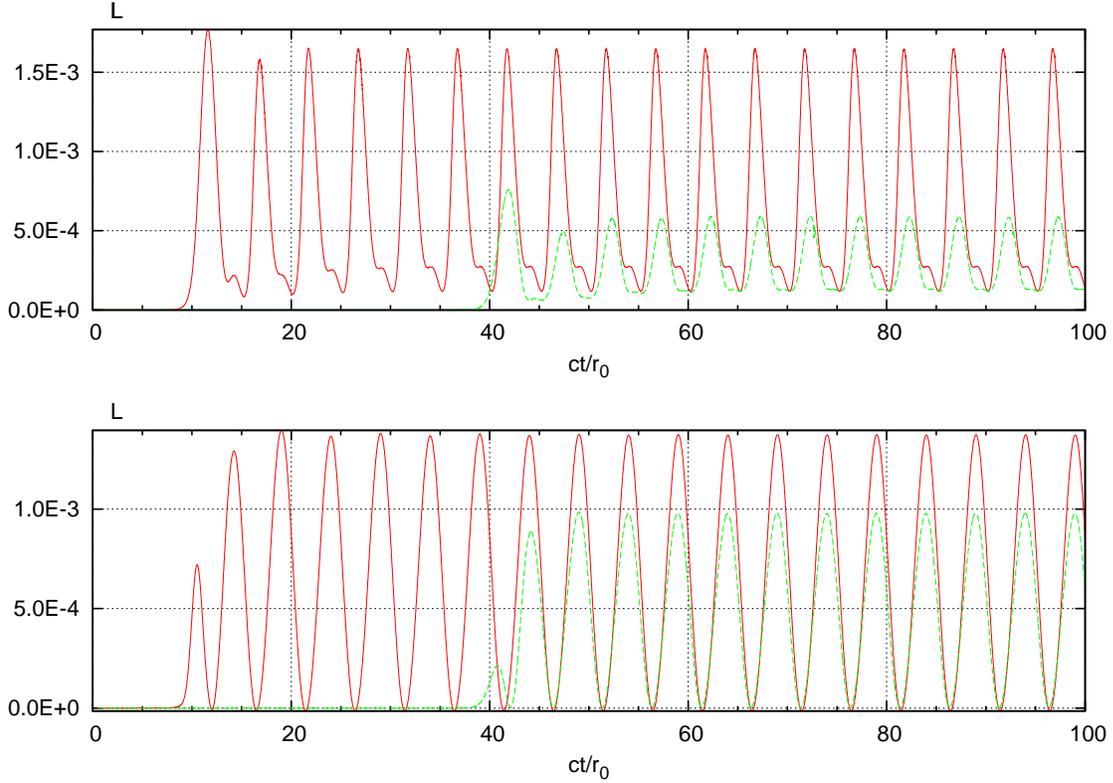}
\caption{
Luminosity as a function of normalized time $ct/r_{0}$.
The top panel shows $ L/(r_{0} B_{s})^{2}c$
for $\sigma_{0}/\nu =5$, and the bottom panel for $\sigma_{0}/\nu = 0.5$.
The two sinusoidal curves correspond to the values integrated at
$r/r_{0} =10$ and $r/r_{0} =40$.
}
\end{figure}

  We examine the dependence of $\sigma _{0}$
on the luminosity in Fig. 2. Time-averaged values integrated 
at the surface corresponding to $ r= \lambda $, $4\lambda $ 
and $8\lambda $ are denoted 
by $L_{1}$, $L_{4}$ and $L_{8}$ ($L_{1} > L_{4} > L_{8}$).
The luminosities irrespective to the position
($r > \lambda$) approach 
$L_{*} \equiv 9.8\times 10^{-4} (r_{0} B_{s})^{2}c $
in the limit of small $\sigma_{0}$, where $L_{*}$ is 
the value of dipole radiation in vacuum,
estimated in Eq. (\ref{eqn.TYP}).
All the values  $L_{k}$ $(k=1,4,8)$ slightly
decrease with increasing $\sigma _{0}$, 
but tend to increase for $\sigma _{0} /\nu  > 2$.
The luminosity minimum is located near $\sigma _{0} /\nu  \sim 2$.
The dependence of $\sigma_{0}$ can be explained by two effects.
One is the Ohmic dissipation:the decay timescale increases with it.
The other is flowing currents produced in our model:the magnitude 
generally increases with $\sigma_{0}$.
In the vacuum limit,  $\sigma \equiv 0$, there are no currents and no 
dissipation of the electromagnetic energy.
In the small region, the energy is quickly dissipated, and the 
reduction of the flux is more evident at great distances.
With increase of $\sigma_{0}$, the Ohmic dissipation  becomes less effective, 
and the amount of flowing currents increases. 
They produce electro-magnetic field structure different from 
that of vacuum in large $\sigma_{0}$ region.
The induced fields dominate, and the outgoing flux 
is independent of the inner boundary condition 
as long as it is estimated outside the induction region.

\begin{figure}[!htb]
%
 \includegraphics[scale=0.75, angle=0]{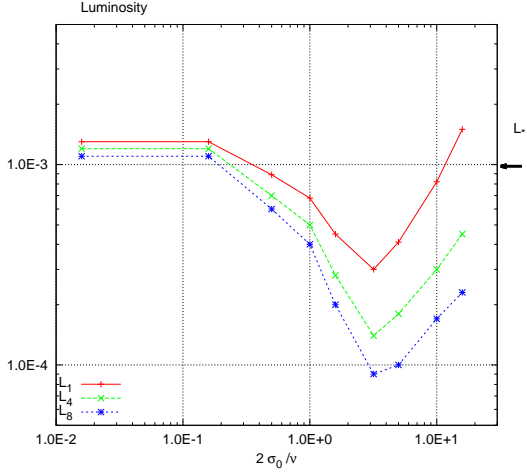} 
\caption{
Time-averaged luminosity $L/(r_{0}B_{s})^2c$
as a function of parameter $2\sigma_{0}/\nu$.
The arrow near the right-hand axis denotes 
the typical value $L_{*}$ by Eq.(\ref{eqn.TYP}).
}
\end{figure}

 Figure 3 shows time evolution of electromagnetic fields.
It is almost periodic, and the evolution during half of the 
oscillation period ($0.5\nu^{-1} = 5r_{0}/c $) is shown.
The top and bottom panels show the results for 
low conductivity ($\sigma _{0} /\nu =0.5$)
and high conductivity ($\sigma _{0} /\nu =5$), respectively.
The contours of the function $S (=B_{\phi} r \sin \theta )$
clearly represents a wave propagating at speed $c$.
The oscillation behavior becomes apparent in the wave zone
$r > \lambda =10r_{0}$. The angular dependence at which the amplitude
is maximum on the equator forms a dipolar radiation pattern. Although 
octupole radiation may be involved, its effects are rather small.
The magnetic flux function $G$ describing poloidal
magnetic fields is also plotted.
The pattern is dipolar in the inner region $r < 10 r_{0}$
but gradually deviates from this as the radius increases. 
This fact becomes much clearer at high conductivity.
The contour of $G$ in Fig. 3 
shows remarkable topological changes, i.e., 
reconnection due to resistivity: a small oval becomes detached
from the outer part ($r \sim 15 r_{0}, \theta \sim \pi/ 2$)
and shrinks beyond that region.
In addition to the interesting behavior of poloidal fields,
the magnitude of $B_{\phi} $ is considerably increased.
Thus, poloidal and toroidal components of the magnetic fields are 
effectively coupled with each other through induced current at 
high conductivity. The induced current disappears at larger radii
in our model ($ \sigma \to 0$), and the magnetic field eventually 
becomes dipolar with some radiative component at infinity.

\begin{figure}[!htb]
%
  \includegraphics[scale=1.5, angle=0]{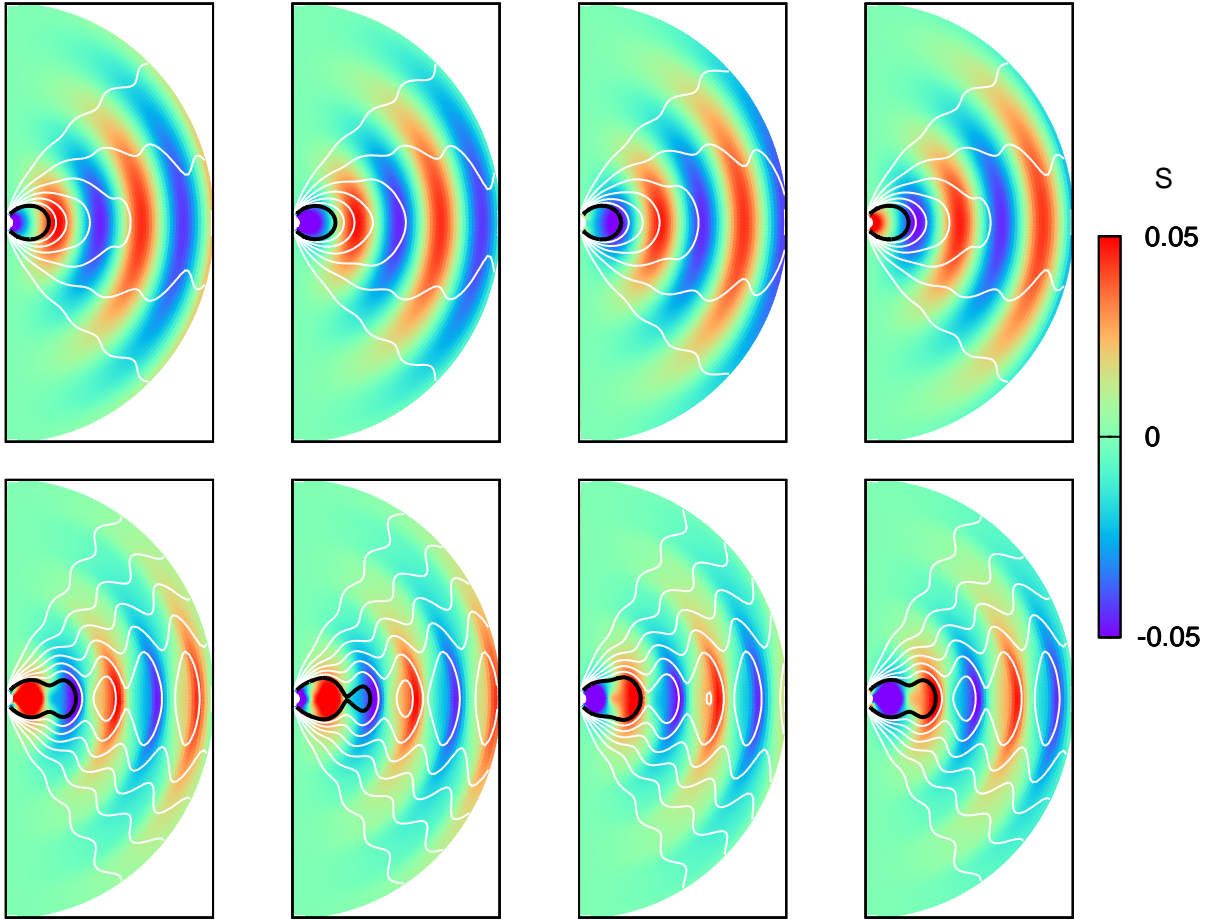} 
\caption{
Time evolution of toroidal and poloidal magnetic fields
in meridional plane ($ 0 \le \theta \le \pi$ and 
$ r \le 25 r_{0}=2.5\lambda$). 
Color contour represents $S (=B_{\phi} r \sin \theta )$
normalized by $B_{s} r_{0}$. Lines represent the level of the 
magnetic function $G/(B _{0} r_{0}^{2}$) 
for $0.03 \times n (n=1,2, \cdots, 7)$.
The innermost contour line is written in black. 
Contours of larger $G$ within it, which are almost unchanged, 
are omitted.
The top and bottom panels show the results for low 
conductivity ($\sigma _{0} /\nu =0.5$),
and high conductivity ($\sigma _{0} /\nu =5$), 
respectively. From left to right, the panels correspond to 
$ ct/r_{0}=$ 50.0, 51.7, 53.3 and 55.0.
}
\end{figure}

 \section{Discussion}
  We studied time variation in a magnetosphere
disturbed by torsional shear oscillation by 
using resistive electrodynamics.
In this formalism, the current is described by Ohm's law and
the conductivity, and the electromagnetic fields are solved 
numerically.
The results indicate that the induced 
current increases with the increase in conductivity, 
the fields are substantially modified, 
and the energy flux is thereby increased. Beyond a certain critical value
of the conductivity $\sigma _{0}/ \nu \sim {\mathcal O}(1)$,
induced components dominate, and the inner boundary condition 
becomes less important. The behavior is however localized 
since the induced current disappears in the outer part. 
The exterior is assumed to approach vacuum in our model, 
in which the conductivity approaches zero.
The extent of energy loss in the form of radiation is almost the same as
that in vacuum.
  It is interesting to compare our results with those for
a pulsar magnetosphere by \citet{2012ApJ...746...60L},
where the conductivity is assumed to be uniform everywhere.
The results in \citet{2012ApJ...746...60L} show that 
the luminosity increases with conductivity 
and approaches that of force-free calculation
in the limit of high conductivity ($\sigma _{0}/ \nu \gg 1 $).
Mathematically, the force-free limit is written as
$ \sigma \to \infty$ and 
$E_{0} \to  0$, but $  \sigma E_{0} $ is finite
in Eq. (\ref{eqn.LST}).
In our numerical simulation, we found that 
$E_{0} $ does not necessarily approach zero everywhere
($E_{0} B_{0} = \vec{E} \cdot \vec{B} \ne 0$).
The electromagnetic wave components $\vec{E}_{w}$ and $\vec{B}_{w}$ 
satisfy $\vec{E}_{w} \cdot \vec{B}_{w} =0$, however, there
is another dipole component $\vec{B}_{d} $,
for which $\vec{E}_{w} \cdot \vec{B}_{d} \ne 0$.
It is not clear 
whether the problem arising in the limit of high conductivity
depends on the spatial distribution model or on the numerical method.

The electric conductivity differs depending on the direction
in the strongly magnetized regime, in which case
the isotropic form of Ohm's law (\ref{eqn.Ohm})
may be replaced by a tensor.
As a brief discussion of directional differences,
the conductivity is assumed to arise from collisions, and
some terms in the generalized Ohm's law are ignored.
The conductivity parallel to the magnetic field is given by
$\sigma _{\parallel}= e^{2} n_{e}/(m_{e} \nu_{c})$, 
and that in perpendicular direction is given by
$\sigma _{\perp} = \sigma _{\parallel}/(1+(\omega_{B}/\nu_{c})^2)$,
where 
$\omega_{B}= eB/(m_{e}c)$ is the magnetic gyro-frequency
and $\nu_{c}$ is the collision frequency
(e.g., \cite{2006ASSL..340.....S}).
Using the Goldreich-Julian number density $ n_{e}= B \nu/(e c)$, 
the ratio is reduced to $\sigma _{\perp}/ \sigma _{\parallel}$
$ = (1+(\sigma _{\parallel}/\nu )^2)^{-1}$.
The isotropic form may be no longer valid in the limit of
$\sigma _{\parallel}/\nu \gg 1$, where
$ \sigma _{\parallel} \gg \sigma _{\perp}$. 
In our numerical simulation, the parameter is 
limited to $\sigma _{0}/\nu  < 10$.
The anisotropic form of Ohm's law would be adequate for larger 
values of to $\sigma _{0}/\nu$.
Thus, conductivity is an important factor, but
its magnitude and spatial distribution
are poorly understood at present.
A more elaborate treatment is necessary for constructing a more 
realistic model.

  The outgoing energy flux is essential for estimating the damping 
mechanism of the oscillations.
The luminosity $L$ calculated in this paper 
is of the same order as that in vacuum, 
although it depends on the conductivity.
Appropriate scaling of the shear oscillation with $l=2$ leads to 
$L \sim 2 \times 10^{37}$
$(\nu/30{\rm Hz})^{4}(\epsilon/10^{-2})^{2}(B/10^{14}{\rm G })^{2}$ 
ergs s$^{-1}$.
The damping timescale for the energy $E \sim 10^{45}$ergs is 
$t_{d}=E/L$ $\sim $ a few years.
This is a rather long period. It is difficult to significantly increase
radiation loss, even if there exists a magnetosphere.
The loss to infinity is not so efficient in the exterior vacuum region.
The oscillation energy slowly escapes into the radiative 
zone, which is located beyond the region corresponding to a few wavelengths.
In the presence of a magnetosphere,
both ingoing and outgoing fluxes are likely
to be produced due to interaction inside that zone,
and current flows also change direction there.
The inward energy flux and return current flows may be non-negligible 
in magnitude and may be 
dissipated as Joule heating in the region with higher resistance.
Thus, the problem is extended to a globally coupled system 
between a neutron star/crust and a magnetosphere,
which is more complex to model.
The dynamics of coupled oscillations is an interesting
topic for future research.
%

A model of a twisted magnetosphere would be more realistic since global 
electric current flows through the stellar 
surface\citep{2002ApJ...574..332T,2007ApJ...657..967B}.
The exterior of a star before flares is modeled by a purely dipole field, 
and in the present model current flows are allowed only during 
the oscillations. However, the effects of twisting may be significant.
An interesting simulation by using a time-dependent
force-free electrodynamics model was recently 
presented in \cite{2012ApJ...754L..12P}, where
the evolution of the twisted magnetosphere was calculated. 
It was found that slow shearing at the surface leads 
to twisting and an abrupt reconnection.
The evolutionary time scale ($\gg 1$ s) of that model is longer than 
that of ours ($< 1$ s).
Such a gradual twisting process would also be an interesting 
topic for future research.

 \section*{Acknowledgements}
This work was supported in part by the Grant-in-Aid for Scientific Research
(No.21540271 and No.25103514) from the 
Japanese Ministry of Education, Culture, Sports, Science and Technology.


  \end{document}